\documentstyle[aps,eqsecnum,tighten,epsf,rotate]{revtex}
\newcommand{\mod}[1]{\vert{#1}\vert}

\newcommand{\disc}{\displaystyle}

\newcommand{\vcr}{V_{C}(r)}
\newcommand{\vsigmatau}{V_{\sigma\tau}(r)}
\newcommand{\vtensor}{V_{T}(r)}

\newcommand{\lhisb}{{\cl}_{{\disc\chi}\mbox{sb}}}

\newcommand{\rv}{\vec{r}}
\newcommand{\vecr}{\vec{r}}
\newcommand{\vecx}{\vec{x}}
\newcommand{\vecrone}{\vec{r_{1}}}
\newcommand{\vecrtwo}{\vec{r_{2}}}
\newcommand{\vecri}{\vec{r_{i}}}
\newcommand{\fpi}{F_{\pi}}
\newcommand{\tauv}{\vec{\tau}}
\newcommand{\tauone}{ \tauv_{1}   }
\newcommand{\tautwo}{ \tauv_{2}   }

\newcommand{\cl}{\cal {L}}

\newcommand{\gzero}{g^{\prime}_o}

\newcommand{\apm}{{\hat{A}}^{+}_{1}{\hat{A}_2}}
\newcommand{\aone}{\disc{\hat{A}_1}}
\newcommand{\atwo}{\disc{\hat{A}_2}}
\newcommand{\uzero}{\disc{U_0}}
\newcommand{\uone}{{\disc}U_{1}}
\newcommand{\utwo}{\disc{U_2}}
\newcommand{\be}{\begin{equation}}
\newcommand{\ee}{\end{equation}}
\newcommand{\ba}{\begin{array}{l}}
\newcommand{\banonum}{\begin{eqnarray*}}
\newcommand{\ea}{\end{array}}
\newcommand{\eanonum}{\end{eqnarray*}}
\newcommand{\banum}{\begin{eqnarray}}
\newcommand{\eanum}{\end{eqnarray}}
\newcommand{\bb}{\begin{thebibliography}}
\newcommand{\eb}{\end{thebibliography}}
\newcommand{\ci}[1]{\cite{#1}}

\newcommand{\bi}[1]{\bibitem{#1}}
\newcommand{\lab}[1]{\label{#1}}
\newcommand{\re}[1]{(\ref{#1})}
\newcommand{\Tr}{\mbox{Tr\,}}
\newcommand{\dd}{\partial}
\newcommand{\edc}{\end{document}}

\newcommand{\ga}{\disc{g_{A}}}
\newcommand{\thprim}{{\Theta}'}

\newcommand{\mpi}{{\disc{m}_{\pi}}}

\newcommand{\gpinn}{\disc{g_{{\pi}NN}}}

\newcommand{\mn}{M_{N}}
\newcommand{\mnnuc} { \mn^{*} }
\newcommand{\md}{M_{\Delta}}
\newcommand{\mh}{M_{H}}

\newcommand{\lamm}{{\disc\lambda_{M}}}
\newcommand{\rhozero}{{\disc\rho_{o}}}

\newcommand{\ltwo}{{\cl}_{\mbox{2}}}
\newcommand{\lfora}{{\cl}_{\mbox{4a}}}

\newcommand{\vnn}{V_{NN}}

\newcommand{\stwo}{{\disc}s_{2}}

\newcommand{\bea}{\begin{eqnarray}}
\newcommand{\bc}{\begin{center}}
\newcommand{\ec}{\end{center}}
\newcommand{\eea}{\end{eqnarray}}

\newcommand{\fpisix}{\disc\frac{\fpiq}{16}}
\newcommand{\vectau}{\vec{\tau}}
\newcommand{\vecpi}{\vec{\pi}}
\newcommand{\fpiq}{F_{\pi}^{2}}
\newcommand{\lpi}{ {\cl}_{{\disc\pi}} }
\newcommand{\lsk}{ {\cl}_{{\disc\mbox{sk}}} }
\newcommand{\leff}{ {\cl}_{{\disc\mbox{eff}}} }
\newcommand{\mpiq}{{\disc{m}_{\pi}^{2}}}
\newcommand{\dmuup}{\dd^{\mu}}
\newcommand{\dmud}{\dd_{\mu}}
\newcommand{\dv}{\vec{\nabla}}
\newcommand{\polar}{\hat{\Pi}}
\newcommand{\uopt}{\hat{U}_{opt}}
\newcommand{\chip}{{\disc{\chi}_{p}}}
\newcommand{\chis}{{\disc{\chi}_{s}}}
\newcommand{\alfap}{{\disc{\alpha}_{p}}}
\newcommand{\alfas}{{\disc{\alpha}_{s}}}

\newcommand{\di}{\dd_{ i } }

\newcommand{\ratga}{\ga^{*}/\ga}
\newcommand{\ds}{\displaystyle}

\newcommand{\doublespace}{
\renewcommand{\baselinestretch}{1.6}\large\normalsize}

\doublespace
\begin{document}
\draft
\title {  \bf  {Medium Modification of Nucleon Properties
in Skyrme Model }}
\author{A.M. Rakhimov \thanks{Permanent address:
Institute of Nuclear Physics, Tashkent, Uzbekistan},
 M.M. Musakhanov$^{\ddagger}$, F.C. Khanna \thanks
{E-mail : khanna@phys.ualberta.ca}
  and    U.T. Yakhshiev$^{\ddagger}$  }

\address{
Physics Department,     University of  Alberta
Edmonton,     Canada  T6G2J1 \\
 and\\
TRIUMF,     4004 Wesbrook Mall, Vancouver,     British Columbia,
Canada,     V6T2A3\\
$^{\ddagger}$  Theoretical  Physics Department ,  Tashkent State University,\\
  Tashkent  ,    Uzbekistan   (CIS).\\}
\maketitle

\begin{abstract}
A Skyrme type Lagrangian for a skyrmion  imbedded in a  baryon
rich environment is proposed. The dependence of static nucleon
properties  and nucleon - nucleon  tensor interaction  on
nuclear density is investigated.

\end{abstract}


\newpage
\large

\section{Introduction}
\indent

Properties of a single nucleon in free space are understood in classical terms
by means of a soliton -like
solution of nonlinear Lagrangians like
 the Skyrme Lagrangian  \ci{anwzb}. Models
have been constructed to
 consider pion - nucleon and nucleon - nucleon ( NN)  interactions
\ci{nn,oursingap,yabu} and even to deform the nucleons \ci{ourdeform}.
An important outstanding problem is  the study of
 these soliton -  like structures
in a many body system. In heavy ion collision the
 properties of hadron are to be
studied in a hot, non - zero temperature,
 and dense, density much larger than
$\rho_o$, nuclear matter (here $\rho_o$ is the density of
 normal nuclear matter).
There
are good reasons to believe that the properties of hadrons like mass, radii
and coupling to external currents change in a hot and dense matter
medium. Then the modification of   NN interaction has to be understood.
 The
medium plays an important role in changing
the strength of the interaction and
the relative strengths of central, tensor  and spin-orbit interactions.

Quantum Chromodynamics, the fundamental
 theory of  strong interactions, has to
be replaced with  an effective theory to consider the
 nuclear interactions in
the medium in any consistent manner.
In arriving at an effective theory  the
important  constraints of QCD, chiral symmetry
 and scale invariance  have to
be retained.
Recent considerations of Brown  and Rho \ci{br} are an attempt to find and
elucidate such a theory. In constructing an effective Lagrangian,  $\leff$,
the in- medium modification is  reflected by a change in  the vacuum
expectation value of a dilatation  field.
The resulting Lagrangian, which obeys the trace anomaly  of QCD
almost coincides with Schechter's Lagrangian \ci{gomm} in form but includes
some modified parameters. However this in- medium $\leff$ does not take into
account the possible modification  of the chiral field, since it is considered
here as a massless Goldstone boson. On the other hand it is quite natural to
assume that,   $\leff$ has to include  the direct distortion of chiral
fields. In fact, in a linear approach, Skyrme Lagrangian describes the free pion
field and its in - medium modified version must be relevant to pion fields in
nuclear matter.

  The aim of this paper  is to
 consider a nucleon placed in a nucleus and try to describe
this nucleon in the framework of the Skyrme model,
taking into account  the
influence of nucleus as a medium (see also \ci{meissner,kalber}).
It is not our goal  to describe the whole nuclear system. Instead we shall
concentrate on the changes of nucleon properties  embedded in nuclei taking
into account the influence of baryon rich environment as an external
parameter. Our basic idea is that,  in the linear approach the $\leff$ should
give the well known \ci{pinbooks}  equation for the pion field
$\dmuup\dmud\vec\pi+(m_{\pi}^2+\polar)\vec\pi=0,$
where  $\polar      $  is the polarization function or the self energy
of pion field in the medium.

The paper is organized as follows: In Sec. II, we propose a modified Skyrme
Lagrangian $\leff$ - including the distortion of chiral field in the medium. The soliton
- like solutions of this Lagrangian represent a skyrmion embedded in the
nuclear medium;  the Lagrangian is applied to calculate the static
properties of the nucleon in Sec. III. In Sec. IV, we consider the
possible modifications of nucleon - nucleon tensor interaction due to the
presence of the medium. We summarize and discuss the results
in Section V.

\section{The in-medium Skyrme Lagrangian}
\setcounter{equation}{0}

\indent
The Skyrme model is a theory of nonlinear meson fields where baryons
 can emerge as soliton solutions.
  The Skyrme Lagrangian may be written as \ci{anwzb}:
\be
\ba
\lsk=\ltwo+\lfora+\lhisb,\\
\quad \\
\ltwo=-\fpisix\Tr({\dv}U)\cdot({\dv}{U}^{+}),\\
\quad \\
{\lfora}={\disc\frac{1}{32{e}^{2}}}{\Tr}[U^+{\di}U, U^{+}{ \dd_{j} }U]^2, \\
\quad \\
\disc \lhisb = - \frac{\fpiq}{16}{\Tr}[(U^{+}-1)\mpiq(U-1)],
\lab{lfreesk}
\ea
\ee
where in usual notations $\fpi=2f_{\pi}$, $e$ - the Skyrme parameter.
The expansion around the vacuum value $(U \approx 1)$
\be
U=\exp[2i(\vectau\vecpi)/\fpi]\approx 1+\frac{2i}{\fpi}(\vectau\vecpi)+
\dots
\lab{uexpan}
\ee
in \re{lfreesk} gives a Lagrangian
for free linear pion field:
\be
{\cal L}_{sk} \approx {\cal L}_{\pi} = \displaystyle
-\frac{1}{2}(\vec\nabla \vec\pi)^2 -\frac{1}{2}m_{\pi}^2\vec\pi^2.
\ee

Let  us consider a skyrmion inserted in a nucleus. It is well known
 that pions
in nuclei are described \ci{pinbooks} by the Lagrangian
\be
{\cal L}_{\pi}^{\star}=\lpi-\frac{1}{2}\vecpi\polar\vecpi
\lab{lpimedium}
\ee
 (the asterisk indicates  the medium )
where ${\hat {\Pi}}$
is the  self energy, or, the polarization operator, which characterizes
the modification of the pion propagator in the medium.
 Bearing
in mind an  expansion like \re{uexpan} we may generalize
\re{lfreesk} as:
\be
\ba
\lsk^{*}=\ltwo+\lfora+\lhisb^{*}\quad ,\\
\quad \\
\lhisb^{*}=\displaystyle
-\frac{F_{\pi}^2\mpiq}{16}\Tr[(U^{+}-1)(1+\polar/{\mpiq})(U-1)]
\lab{lskmedium}
\ea
\ee
with the only modified term,  $\lhisb^{*} $,
which  describes  the distortion of pion
field  in the medium.

The calculation of the pion self energy in the coordinate space
within the Skyrme model in a self consistent way is a special problem.
It is not our goal to calculate it in the present paper,
since we are not describing
the whole system of nucleons in the framework of the Skyrme model.
Instead,  in the coordinate space we use a simple relation between
$\polar$ and the pion-nuclear optical potential  $\uopt$ :
$\polar \approx 2\omega_{\pi}\uopt$ \ci{pinbooks}.

In general, the operator $\hat\Pi$ acts on
 $\vec{R}$ -coordinate of center of mass of soliton as well as on its
internal collective coordinate  $\vec{r}$ i.e.
 $\hat\Pi=\hat\Pi(\vec{R}, \vec{r}-\vec{R})$
\footnote{We are indebted to the referee who
 drawn our attention to this point.}.
 For heavy nuclei the $R$ dependence is
weak and for homogenous nucleus it may be
neglected totally.
So, letting   $\hat\Pi=2\omega_{\pi}\uopt(\vec{r})$ in Eq. \re{lskmedium}
we may choose an optical potential widely used
in the literature.
It is clear  from \re{lskmedium} that  when $\uopt$ is a local one, like
"laplacian potential"   \ci{pinbooks} the modification of the Lagrangian is
trivial
and mainly consists  in   changing the  pion mass into an effective mass
$m_{\pi}^{*} =\mpi\sqrt{1+2\uopt/\mpi}$ in the medium.

Clearly, the most interesting case is to use
the nonlocal Kisslinger potential,
 used both in describing   pionic atoms and pion nuclear
scattering. At   threshold, when $\omega_{\pi} \approx \mpi$,
it may be represented
in a  schematic way   \ci{pinbooks}:
\be
\polar=\chis(r)+\dv\cdot\chip(r)\dv,
\lab{polarschem}
\ee
where $\chis$ and $\chip$ are some functionals of S- wave and P- wave    pion
nucleon
 scattering lengths, and   nuclear density - $\rho(r)$.
Using \re{polarschem} in \re{lskmedium} and bearing in
 mind integration by
part
 we obtain the following Lagrangian:
\be
\ba
\lsk^{*}=\ltwo^{*}+\lfora+\lhisb^{*},\\
\quad \\
\ltwo^{*}=-\fpisix\alfap(r)\Tr({\dv}U)\cdot({\dv}{U}^{+}),\\
\quad \\
{\lfora}={\disc\frac{1}{32{e}^{2}}}{\Tr}[U^+{\di}U, U^{+}{ \dd_{j} }U]^2, \\
\quad \\
\lhisb^{*}=\displaystyle \frac{F_{\pi}^2\mpiq}{16}\alfas(r)
\Tr(U+U^{+}-2),
\lab{lfinal}
\ea
\ee
where $\alfap(r)=1-\chip(r)$,
 $\chip(r)$ - pion dipole  susceptibility of the medium, and
$\alfas(r)=1+\chis(r)/\mpiq$.

Thus, the nonlocal Kisslinger potential modifies not
only the pion mass term but also the kinetic term - $\ltwo$.
 Note
that in our model the fourth  order derivative term -  $\lfora$ remains
unchanged. This is not surprising, since this term
 corresponds to the infinite mass limit of the $\rho$ - meson term
\ci{bhaduri}, whose self energy operator is not considered
here.  Thus  our basic Lagrangian
is given in Eq. \re{lfinal} and  will be used
to  investigate the modification of nucleon properties in  the medium.

\section{The in-medium nucleon properties}
\setcounter{equation}{0}

\indent

In general, the Lagrangian in Eq. \re{lskmedium}
is conceivably valid both for finite and infinite nuclei.
However in practical calculations in finite nuclei there may arise
some difficulties concerned with surface effects and localization
of the Skyrmion in nuclear medium. For the simplicity
we'll consider medium modifications in homogeneous nuclear matter. In this
case $\chip(r)$ and $\chis(r)$ in the Lagrangian are
clearly constants ($\chip(r)\equiv\chip, \chis(r)\equiv\chis)$
and the skyrmion may be assumed to have spherical symmetry.

For the spherically symmetric static \quad Skyrme ansatz,
\quad$U(r)=U_0=\exp{ (i\vec{\tau}\hat{r}\Theta(r))}$,
  $ \hat{r}=\vec{r}/\mod{r}$,
the mass functional for the dimensionless $x=e{\fpi}r$ has  the form:
\footnote{Here  the skyrmion is assumed to be placed
 right in the center of mass of nucleus.}
\be
\ba
\mh^{*}= {\disc\frac{4{\pi}{\fpi}}{e}}{\disc
\int\limits_0^{\infty}dx}(\tilde{M}_{2}^{*}+
\tilde{M}_{\mbox{4a}}+
\tilde{M}_{{\disc\chi}\mbox{sb}}^{*}),  \\
\quad \\
\tilde{M}_{2}^{*}=
{(\thprim^{2}x^{2}/2+s^{2})}(1-\chip)/4, \\
\quad \\
\tilde{M}_{\mbox{4a}}=s^{2}(d/2+\thprim^{2}), \\
\quad \\
\tilde{M}_{{\disc\chi}\mbox{sb}}^{*}=
(1-c)x^{2}{\beta}^{2}(1+\chis/\mpiq)/4,
\lab{mass}
\ea
\ee
where
 $c\equiv\cos(\Theta)$, $s\equiv\sin(\Theta)$,
$d={(s/x)}^{2}$,  ${\beta}={\mpi}/({\fpi}e)$.
Since the  nuclear dipole susceptibility, $\chip$, is
nearly proportional
to the nuclear density $\rho$, for  large densities the
 $\tilde M_2^*$ term,  arising from $\ltwo^{*}$
 becomes negative and a skyrmion may disappear. Let's
discuss this point in detail. The Euler - Lagrange equation for the
shape function,  $\Theta (x)$, is given as:
\be
\ba
\Theta^{''}[{\disc}x^{2}\alfap+8s^{2}]+
2{\disc}{\thprim}x\alfap+4\thprim^{2}\stwo
-[\stwo\alfap+4d\stwo+x^{2}{\beta^2}s\alfas]=0,
\lab{urav}
\ea
\ee
where $\stwo \equiv \sin (2\Theta)$ and a prime corresponds to a derivative
with respect to $x$.
As  we are not interested  in describing the nuclear
system as a whole, we use  solutions of    Eq.  \re{urav} with $\Theta (0)=\pi
$ corresponding to the baryon number  $B=1.$
 The asymptotic behavior of $\Theta (x)$ at large distances
is similar to that  for the free case:
\be
\ba
\displaystyle\lim_{x\to\infty}\Theta(x)=
\gamma\disc\frac { (1+{\beta}^{*}x)\exp{ (-{\beta}^{*}x)} } {x^2},\\
\quad \\
\beta^{*}=\beta\disc\sqrt\frac
{1+\chis/\mpiq}{1-\chip}.
\lab{asymp}
\ea
\ee
It is well known \ci{pinbooks} that for finite   nuclei the pion
susceptibility is always less than unity, $\chip < 1$. However,
for infinite nuclear matter with a constant
 density $\rho = \lambda \rhozero$
  $(\rhozero = 0.5 m_{\pi}^{3})$     there is some critical value of
$\lambda$
when the expression under the square root sign in
Eq. \re{asymp} becomes negative which leads to an exponentional dissipation of
soliton solutions. Thus the condition for survival  of a skyrmion in the dense
matter is equivalent to  comparing the  dipole susceptibility with  unity as in
the usual pion nuclear physics  \ci{pinbooks}. This result may be compared with
the model proposed in
Ref. \ci{mishustin}, where there are  no skyrmion
 solutions even for  real nuclei.

In order to carry out numerical calculations we adopt the following expressions
for
 $\chis$ and  $\chip$ \ci{pinbooks}:
\be
\ba
\chis=-4\pi{\eta}b_o\rho,
\quad
\chip=\disc\frac{\kappa}{1+\gzero\kappa},
\quad
\kappa=4{\pi}c_o\rho/\eta,
\lab{chips}
\ea
\ee
where $\eta = 1 + \mpi/\mn$ - a kinematical factor, $\mn$ - mass of
the nucleon.
The parameters $b_o, c_o $ are some
effective pion - nucleon S and P wave scattering lengths, and $\gzero$ -
Lorentz-Lorenz    or correlation parameter.

We use the following set of empirical parameters
$b_o = - 0.024 m_{\pi}^{-1}$, $c_o =  0.21m_{\pi}^{-3}$
\ci{tausher}.
Parameters $\fpi$ and $e$ have  the values
$\fpi = 108 MeV$, $e = 4.84$ as in  Ref.\ci{anwzb},
 so for the free nucleon $\mn = 939 MeV$ and $\md=1232MeV$.
Using these  values in \re{asymp} and \re{chips}
the critical density of nuclear matter $\rho_{crit}$
may be estimated, when a stable solution of     Eq.  \re{urav} (that is a
skyrmion) does  not
exist as  $\rho_{crit} \geq 1.3 \rhozero  $ and
$\rho_{crit} \geq 3 \rhozero$
 ($ \rho_o=0.5m_{\pi}^3$-normal nuclear density)  for $\gzero =1/3$ and
  $\gzero = 0.7$  respectively.
 Clearly  for  a real nuclear when  $\rho \leq  \rhozero$
 this  model is valid.
 Standard
canonical quantization method \ci{anwzb}
gives the familiar expressions for mass
of the nucleon  and $\Delta$ - isobar
\be
\ba
\mn^{*}=\mh^{*}+3/8\lambda_M^{\star},\\
\quad \\
\md^{*}=\mh^{*}+15/8\lambda_M^{\star},
\lab{mn}
\ea
\ee
 where $\mh^{*}$
- soliton mass \re{mass}, $\lambda_M^{\star}$ is the moment of inertia of
 the rotating skyrmion:
\be
\ba
\lambda_M^{\star}={\disc\frac{8{\pi}}{3{e}^{3}{\fpi}}}{\disc
\int\limits_0^{\infty}dx}x^{2}s^{2}[1/4+
\thprim^{2}+d ]
\ea
\lab{lamm}
\ee
where $\Theta $ is the solution of      Eq. \re{urav}.
The mass  $\mh^{*}$ may be interpreted as a mass of a
soliton of the nonlinear pion fields affected by the medium.
Note that, the moment of inertia $\lambda_M^{\star}$ does not include
  the nuclear density $\rho$  explicitly.
 The reason is that the nonstatic parts of the self energy operator
 are  not included in the calculations.
 Similarly, the isoscalar and isovector
mean
square radii, defined by zero components of  baryon  and vector  currents,
have the same formal expressions as in the free case:
\be
\ba
<r^2>^*_{I=0}=-\ds\frac{2}{e^2F_{\pi}^2\pi}
\int\limits_0^{\infty}x^2\Theta^{\prime}s^2dx,\\
\quad \\
<r^2>^*_{I=1}=\ds\frac{1}{e^2F_{\pi}^2}
\frac{\ds\int\limits_0^{\infty}x^4 s^2[1+4(\Theta^{\prime 2}+d)]dx}{
\ds\int\limits_0^{\infty}x^2s^2[1+4(\Theta^{\prime 2}+d)]dx}.
\lab{radii}
\ea
\ee
Changes in the moment of inertia and size of
the nucleon are not crucial, since they are caused only by a modification of
the profile function $\Theta$.
In contrast, the expression for isovector magnetic moments defined by
the space component of the vector current:
\be
\displaystyle{\mu_{I=1}}=\frac{1}{2}{\int}
d\vecr\;\;\; \vecr{\times}\vec{V}_{3}
\ee
includes medium characteristics explicitly, which arise  from the
contributions
of the kinetic term $\ltwo^*$ to the vector current
\be
\ba
\vec{V}_{k}=-i\disc{\frac{\fpiq}{16}} (1-\chip)\Tr\tauv(L_{k}+R_{k}) +
 { \disc\frac{i}{16 {e}^{2} }  }  \Tr\tauv\{[L_\nu[L_k ,
L_\nu]]+[R_\nu[R_k,R_\nu]]\}\\
\ea
\lab{current}
\ee
where  $L_{\mu}=U^{+}{\dmud}U, $ $R_{\mu}=U{\dmud}U^{+}$.
Hence for the nucleon in  nuclei  simple relations between
magnetic moments and momentum of inertia
 such as $\disc\mu_{I=1}^{p} = \lamm/3 $  shown  in Ref. \ci{anwzb}
 do not  work.
Table \ref{tab1}
illustrates  modifications  of the static properties of the
nucleon in
infinite nuclear matter.
Here the classical value of the correlation factor $\gzero = 1/3$
is  used.

Early arguments
\ci{celenza}
about changes of the
nucleon size in the medium were, in part,
based on the expectation that
$r^{*}/r = \mn/\mnnuc$, where $r^*$ and $\mnnuc$
 are the nucleon radius and mass
within the nuclear medium,  and $r$, $\mn$ are the same two quantities
for a free nucleon. As it is clear  from Table \ref{tab1},  in the present
model
the renormalization  of the nucleon mass
is much  larger than the renormalization  of
the nucleon radius.
The renormalization of
the nucleon radii in Eq. \re{radii} has been  caused only by a  modification
 of the profile function
$\Theta(r)$  (see Fig.1)  in the nucleus, while the modification of $\mn$ is
caused in addition by the  factor $(1-\chip)$ in  Eq. \re{mass}.

There are no direct experimental values of  static properties
of a nucleon bound in nuclei.
 In contrast  many theoretical approaches   are
proposed to estimate them. Many of them deal with an explanation of
 the EMC effect.  For example, in the nuclear binding model \ci{noble},
$\mnnuc=700 MeV$ (appropriate for $Fe$)
and $\mnnuc=600 MeV$ (appropriate for $Au$)
have  been found. On the other hand a calculation
of the nucleon effective mass $\mnnuc$ is an important  problem in  quantum
hadrodynamics
(QHD). The recent results obtained by including $\pi, \rho, \omega$ - meson
fields explicitly in the Lagrangian of QHD give $\mnnuc \approx 620 MeV$
at zero temperature for $\rho = \rhozero$ \ci{song}.
In comparison with these results, our model gives
$\mnnuc= 572 MeV$ for normal nuclear matter for $\gzero=1/3$.
Note that, our results are very sensitive to the value of
Lorentz-Lorenz parameter $\gzero $. For example, using
 another
value,  $\gzero=0.4$,  one may get
$\mnnuc=596 MeV$.
This fact is illustrated in Fig.2a,  where the dependence of $\mnnuc$
on   $\gzero$ is plotted for $\rho = 0$,
$\rho = 0.5\rhozero$ and  $\rho = \rhozero$ using
solid, dotted and dashed  curves respectively.
The present approach  is  similar  in some sense to the  soliton model
of Ref. \ci{jandel} where the mean field
approximation for Friedberg-Lee approach is  used.
 A swelling of the nucleon size $\sim 30 \%$
predicted there, is in  good agreement with our result (see Table \ref{tab1}).
On the other hand there is a pion excess model proposed by M. Ericson
\ci{ericsonm},
which explains the swelling by  a distortion of the pion cloud in the medium.
However, she obtained a very large modification of the nucleon size, i.e.
nearly  doubling of the free value of the r.m.s. radius.
In addition in the pion excess model the effect of
 swelling concerns only the isovector   radius,
whereas  in this approach the swelling includes
 both isovector and isoscalar
radii.

Another  interesting phenomenon of pion nuclear physics is that,\quad in
nuclei the axial coupling constant $\ga$, governing Gamow - Teller
transitions,  may be modified significantly from its  free-space
 value $\ga \approx 1.25$. It is  shown
 that $\ga$  is  systematically
renormalized downward in finite nuclei \ci{rho}.
 A most remarkable observations, made in Ref. \ci{buck}
 based on a model independent analysis
of \quad $\beta$ - decay and magnetic moment data of the mirror nuclei,
\quad ($3\leq{A}\leq39$), is that the axial coupling constant
 in nuclei equals   unity
to a very good accuracy: $g_A^* = 1.00 \pm 0.02$, that is
 $g_A^*/\ga=0.8$ for nuclear matter.

Although the Skyrme model, especially in its original version,
gives an  underestimate for the  value of  $\ga$ ($\ga = 0.65$
 for the free case)
we may try to investigate the quenching phenomenon within
the present approach.
It is  easy to show that  the expression for $\ga$ is the
same as in the free case but there is an additional factor in the term arising
from the kinetic term:
 \be
\ba
 \ga^{*}=-\frac{\pi}{3e^2} {\int\limits_{0}^{\infty}}dx\:x^{2}(
g^{*}_{2}(x)+g_{4}(x) ),\\
\quad \\
g^{*}_{2}(x)=(1-\chip) \cdot (\thprim+ \stwo/x),\\
\quad \\
g_{4}(x)=4[\stwo(\thprim^2+d)/x+2{\thprim}d].
\lab{ga}
\ea
\ee
In nuclei ($\chip <1 $)  $\ga$  decreases due to the factor $(1-\chip)$
under the integral in Eq. \re{ga}.
The  decrease of $\ga$ reaches  $38\%$ for nuclear matter ($\rho =\rhozero$
with $\gzero=1/3)$ (Table \ref{tab1}). This is  consistent with the estimates
carried out in
$\Delta$ - hole coupling model using the random
phase approximation:
$\ga^{*}/\ga\approx 0.67 $   for $\rho/\rhozero=1 $ and
$\ga^{*}/\ga\approx 0.8 $   for $\rho/\rhozero=1/2. $
The $\gzero$ dependence of $\ga^{*}/\ga$ is shown in  Fig.2b.
This dependence is in qualitative agreement
with the formula $\ga^{*}/\ga=[1-4{\gzero}L(0)/9]^{-1}$
presented in the review article \ci{rho}.

For the nuclear with constant density
the Lagrangian in  Eqs. \re {lfinal} with a
 representation of the polarization
operator in  Eqs. \re {polarschem}, \re{chips}  has only $2$ parameters
 dictated by pion nucleon
scattering.
Here the effective pion - nucleon   scattering lengths have been used.
 However  in nuclear matter the pion field is
 localized very close to nucleons
in contrast with the case of pionic atoms. One may ask if the present  model is
able to make  predictions about effective scattering lengths  $b_o  $ and
volumes $c_o $ in nuclear matter?
To do this we have to compare our results with  experimental data. The ratio,
$\ratga$, is  well established to be $ 0.8$, while the
 pionic data analysis yields a value of  $ 0.62$ (see the last line of
 Table \ref{tab1}).
In the nuclear matter $c_o$ is reduced by a factor of $2$
almost  independent of the
value of $\gzero$ to get the correct quenching (Table \ref{tab2}).
For this  optimal case the effective nucleon mass is also close to the common
value of $ 700 $ MeV. In addition
 $\ratga$ is not sensitive to the    S -  wave scattering length $b_o$.
A reduction of the effective  P -  wave scattering length in nuclei may be
clearly understood by the  fact that  quenching of  $\ga$ is equivalent to a
reduction of the pion - nucleon coupling constant $\gpinn$ and hence the pion
- nucleon amplitude.

\section{In - medium NN tensor interaction}
\setcounter{equation}{0}

\indent

Not only the static properties of hadrons but also
the dynamical ones are modified by   the presence of the
medium. In - medium NN interaction differs from the corresponding one
in  free space due to  Pauli blocking (which is not considered here)
and due to the modification of propagators of exchanged  mesons
\ci{limachelit}.
 We investigate the nucleon - nucleon interaction potential
  by using the product approximation:
\be
\ba
U (\vecx;\vecrone , \aone;\vecrtwo , \atwo)=
\aone{\uzero}(\vecx-\vecrone)\apm{\uzero}(\vecx-\vecrtwo){\atwo^{+}}\equiv
\uone\utwo, \\
\lab{u1u2}
\ea
\ee
where $\uzero(\vecx-\vecri)$ for $i=1,2$ is the hedgehog
solution $( U_{0}(\vecr)=\exp{ (i\vec{\tau}\hat{r}\Theta(r))}, $
  $ \hat{r}=\vec{r}/\mod{r})$
 located at $\vecri$,  and $\disc{A_i}$  is
 the collective coordinate to describe the rotation.
The  in - medium NN interaction may be  defined by:
\be
\vnn(\rv)=-{\int}d\vecx[{\lsk^{*}}(\uone\utwo )-{\lsk^{*}}(\uone  )
-{\lsk^{*}}(\utwo ) ],
\lab{potencommon}
\ee
where $\vecr $ is the relative coordinate between
two skyrmions
$(\vecr=\vecrone-\vecrtwo).$
The static NN
potential may be obtained by using a standard technique \ci{nn}
which gives the following general representation:
\be
\vnn(\rv)=\vcr+(\tauone\tautwo)(\vec{\sigma}_{1}\vec{\sigma}_{2})\vsigmatau+
 (\tauone\tautwo)S_{12}\vtensor
\lab{vcommon}
\ee

Unfortunately,
the original Skyrme model for the free case cannot describe the intermediate
range attraction in the central potential within this approximation
 \ci{nn}. This  may be  improved by the inclusion of a scalar
$\sigma$ - meson in the Lagrangian \ci{oursingap,yabu},
which is not taken into
account in the present  calculations.
Here, it is more interesting for us
 to consider
the  tensor part,  $\vtensor$,  of $\vnn$ in Eq. \re{vcommon}, caused
 mainly by the exchange of pions,
modified in the medium.
This  part of $\vnn$ plays an important role in spin -isospin excitations
and pion-like excited states in nuclear physics.

Actually for finite nucleus the product ansatz  \re{u1u2}
should be modified taking into account nonspherical effects. But
for homogenous nuclear matter it is as valid as in the case of
free space especially at intermediate and large separations.
In fact,  formally in this approach,
 the main difference between the in - medium case ($\rho\ne0$) and the free one
($\rho=0$) is that the contribution
to the potential arising  from $\ltwo$ and $\lhisb$    should be
multiplied by factors of $(1-\chip)$ and $(1+\chis)$  respectively.
The resulting $\vtensor$ is  presented  in  Fig.3      for normal
nuclear  matter densities  $\rho=0$, $\rho=0.5\rho_o$
and $\rho=\rho_o$
(solid,  dotted and dashed curves
respectively). The parameters of the optical potential in
 Eq. \re{chips} were
chosen  so as to reproduce the relation
$\ratga=0.8$ for nuclear matter
 (II-line of Table \ref{tab2}).
The nucleon - nucleon tensor interaction in a nucleus   appears to be weaker
than it  is in free space ($\rho=0$). This suppression of $\vtensor$ in nuclear
matter was shown by Hosaka and Toki \ci{hosakatoki} using normalized exchange
meson masses in accordance with the   scaling model of Brown and Rho \ci{br}.
For finite  nuclei it was  shown in Ref. \ci{zamikannphys} by analyzing the energy
difference of $T=1$ and $T=0$, $J=0$ states in $^{16}O$.

\section{Discussion and Summary}
\setcounter{equation}{0}

\indent

We have proposed a modified Lagrangian $\lsk^{*}$
 for a skyrmion placed in a nuclear medium.
In constructing the Lagrangian we required that in the linear
approach   \re{uexpan}  it would yield the well known equation
for the pion field:
$\dmuup\dmud\vec\pi+(m_{\pi}^2+\polar)\vec\pi=0.$
Having been satisfied by inclusion of the pion self energy $\polar$
in to the free space Skyrme Lagrangian, this requirement
determined  the explicit coordinate dependence
of $\polar$ in Eq. \re{lskmedium}. Actually, for a moving Skyrmion
with $U=U(\vec{r}-\vec{R}) $ the similar dependence
$\polar=\polar(\vec{r}-\vec{R}) $ should be fixed. Otherwise
the above equation would not be consistent with the medium modified
Lagrangian in Eq. \re{lskmedium}.

 A much more general choice
as   $\polar=\polar(\vec{R},\vec{r}-\vec{R}) $, which is
essential for a finite nuclear, should have given a chance
to get an information about  energy levels of the bound
 skyrmion as well as about in - medium  modification
of its internal parameters (mass, size etc.)
Since the latter is one of the most exiting topics of
nuclear physics, especially on the light of forthcoming
ultrarelativistic heavy - ion experiments  (e.g. at RHIC),
we restricted  ourselves with the study
 of in medium changes in homogeneous nuclear matter,
where the coordinate dependence of $\polar$ is simpler.

As an  input  data, apart from $\fpi  $ and  $  e $, the present
approach
uses the nuclear density and effective pion - nucleon scattering lengths.
In baryon number one sector, $B=1$, the in - medium
nucleon properties can  be
estimated. Within the present modified Skyrme
model  it is also easy to study  the in - medium
nucleon - nucleon interaction
by  using standard product approximation in the sector with $B=2$.
Consideration of other sectors with $B>2$
has no sense, since the description of even light nuclear in
framework of Skyrme model is a long standing problem.

Let us recall here our main results:
\begin{description}
\item[(i)]
 The critical nuclear density $\rho_{crit}$,
 where a skyrmion, hence, a nucleus remains stable is
$\rho_{crit} \leq 1.3 \rhozero  $
and $\rho_{crit} \leq 3 \rhozero$  for $\gzero =1/3$ and
 for $\gzero = 0.7$ respectively.
This fact shows the strong dependence of $\rho_{crit}$
on the Landau parameter
$\gzero$.
\item[(ii)]
The in - medium effects such as the swelling of a nucleon and decrease of its
mass are not as large as  predicted by pion excess models. The in - medium
change of nucleon mass occurs mainly due to the modification of the second
derivative term $\ltwo$ and depends on the size of $c_o$ - isoscalar P- wave
pion nucleon   scattering volume.
\item[(iii)] A  study of the quenching effect of axial
coupling constant, $\ga$,  in the nuclear matter showed that effective  $c_o$
is much smaller than that predicted by the pionic atom analysis.
\item[(iv)]
These modifications can  occur  naturally in  the NN interaction. Particularly
the tensor part of the interaction in  nuclei appears to be weaker than in free
space.
\end{description}

Modification of nucleon properties found in present paper
are understood by means of medium effects on the chiral
nonlinear field and consequently on the shape and mass of the soliton.

Another explanation of the in - medium modifications,
based on scale invariant arguments, has been recently proposed
by Brown and Rho \ci{br}. The authors implemented
the (broken) scale invariance of QCD in the Skyrme model
and suggested that changes in hadron properties
might arise from a universal scaling related to the scaling anomaly of
QCD. However, further analyses \ci{birse} have shown that
these changes must be small due to the large mass of dilaton,
associated here with a glueball.

In a more fundamental level the origin of these changes
is hidden in a partial restoration of chiral symmetry i.e.
in decrease of quark condensate in nuclear matter \ci{br,birse}.
Unfortunately,
 there are  no quark degrees of freedom in the Skyrme model.
So, in the framework of this model it is natural
 to believe that
the modification of nucleon properties in the medium are caused
by the influence of the latter to the nonlinear pion fields.
This influence can be taken into account , for example,
in terms of pion self energy and, in general, would not be covered
only by   a  trivial scale   renormalization of parameters of the
model.

\section*{Acknowledgments}

We thank   M. Birse, A. Mann, V. Petrov and M. Rho  for useful discussions.
M.M.  Musakhanov and  A.M. Rakhimov            are
indebted to the  University of Alberta for hospitality
 during their stay, where   the main part of
this work was  performed. The research of F. Khanna is
 supported in part by
National Science and Engineering Research Council of Canada.

\bb{99}
\bi{anwzb} G.S. Adkins, C.R. Nappi and E. Witten,
       Nucl. Phys. {\bf B228}, 552 (1983);
 G.S. Adkins and C.R. Nappi,
        Nucl. Phys.  {\bf B233}, 109 (1984);
I. Zahed and G.E.Brown,  Phys. Rep. {\bf 142}, 1 (1986).
\bi{nn} T. Otofuji, S. Saito, M. Yasuno, T. Kurihara, H. Kanada, Phys. Rev.  C
 {\bf 34}, 1559 (1986);
       E. M. Nyman and D. O. Riska,
        Phys. Scr. {\bf 34}, 533 (1986);
         A. De Pace, H. M{\"u}ther and A. Faessler,
         Z. Phys. {\bf A325}, 229 (1986);
        H. Yabu and K. Ando, Progr. Theor. Phys. {\bf 74}, 750 (1985).
\bi{oursingap} M. M. Musakhanov and A. Rakhimov,
 Mod. Phys. Lett. {\bf A10}, 2297 (1995).
\bi{yabu} H. Yabu, B. Schwesinger and G. Holzwarth,  Phys. Lett.
  B {\bf 224}, 25 (1989).
\bi{ourdeform} A. Rakhimov,    T. Okazaki,     M.M.  Musakhanov and F.C.
Khanna,
Phys. Lett. B {\bf 378}, 12 (1996).
\bi{br}  G.E. Brown and M. Rho,  Phys. Rev. Lett. {\bf   66}, 2720 (1991);
Phys. Rep.  {\bf 269},  333 (1996).
\bi{gomm} H.Gomm, P.Jain, R. Johnson and J. Schechter, Phys. Rev. D {\bf 33},
3476 (1986).
\bi{meissner}Ulf - G. Meissner, Nucl. Phys. {\bf A503}, 801 (1989).
\bi{kalber} G. Kalbermann, Nucl. Phys. {\bf A612}, 359 (1997).
\bi{pinbooks}T. Ericson and W. Weise, {\it Pions and nuclei},
(Claredon-Press, Oxford, 1988);
J. M. Eisenberg and D. S. Koltun,
{\it Theory of meson interactions with nuclei}, (A Wiley-Interscience
publication, 1980).
\bi{bhaduri} Rajat K. Bhaduri, {\it Models
 of nucleon from quarks to solitons},
 (Addison-Wiley publishing company INC 1988).
\bi{mishustin} E. Mishustin,  Sov. Phys. JETP  {\bf 71}, 21 (1990).
\bi{tausher} L. Tausher, {\it Physics of exotic atoms}, (Erice, 1977,
Frascati
 INFN, 1977).
\bi{celenza} K. S. Celenza, A. Rozental and C. M. Shakin,
  Phys. Rev. Lett. {\bf 53}, 892 (1984).
\bi{noble} J. V. Nobel,  Nucl. Phys. {\bf A329}, 354 (1979).
\bi{song} Song Gao, Yi-Jun Zhang and Ru - Keng Su, Nucl. Phys.
{\bf A593}, 362  (1995).
\bi{jandel}M. Jandel and  G. Peters,  Phys. Rev. D {\bf 30}, 1117 (1984).
\bi{ericsonm} M. Ericson,  Progr. Theor. Phys. Suppl. {\bf 91}, 244 (1987).
\bi{rho} M. Rho,  Ann. Rev. Nucl. Sci. {\bf 34}, 531 (1984).
\bi{buck} B. Buck and  S. M. Perez,  Phys. Rev. Lett. {\bf 50}, 1975 (1983).
\bi{limachelit} G.Q.Li and R. Machleidt,  Phys. Rev. C {\bf 48}, 1702 (1993).
\bi{hosakatoki}A. Hosaka and H. Toki,  Nucl. Phys. {\bf A529}, 429 (1991).
\bi{zamikannphys}D.C. Zheng, L. Zamick and H. Muther,
 Ann. Phys. (N.Y.)  {\bf 230}, 118 (1994).
\bi{birse} M. Birse J. Phys.G: Nucl. Part. Phys. {\bf 20}, 1537 (1994)
\eb

\newpage
\section*{tables}

\normalsize
\begin{table}
\caption{\large
Ratio of the static properties
of the nucleon in the medium (denoted by asterics)
to that of the free nucleon for various values of the
nuclear density $\displaystyle \rho=\lambda \cdot 0.5m_{\pi}^3$
($g^{\prime}_{o} = 1/3$).}
\vskip 0.2cm
\begin{center}
\begin{tabular}{cccccccc}
 &&&&&&& \\
$\lambda$ &
$\ds \frac{M_N^*}{M_N}$&
$\ds \frac{g_A^*}{g_A}$&
$\ds \sqrt{\frac{<r^2>_{M,I=0}^*}{<r^2>_{M,I=0}}}$&
$\ds \sqrt{\frac{<r^2>_{I=0}^*}{<r^2>_{I=0}}}$&
$\ds \sqrt{\frac{<r^2>_{I=1}^*}{<r^2>_{I=1}}}$&
$\ds \sqrt{\frac{<r^2>^*_p}{<r^2>_p} }$&
$\ds \sqrt{\frac{<r^2>^*_n}{<r^2>_n}}$\\
 &&&&&&& \\
\hline
   0.00 &  1.000 &  1.000 &   1.000 & 1.000 &  1.000 &   1.000 & 1.000 \\
   0.50 &  0.788 &  0.845 &   1.121 & 1.170 &  1.060 &   1.095 & 0.969 \\
   0.75 &  0.696 &  0.743 &   1.177 & 1.255 &  1.086 &   1.140 & 0.939 \\
   1.00 &  0.609 &  0.619 &   1.230 & 1.344 &  1.113 &   1.188 & 0.900 \\
\end{tabular}
\end{center}
\label{tab1}
\end{table}
\begin{table}
\caption{\large
Nucleon effective  mass -$M_N^* $ and   modification of the nucleon
size in normal nuclear matter ($\rho=\rho_o=0.5m_{\pi}^{3}$).
The effective  pion- nucleon  scattering lengt -$b_o$  and
scattering volume -$c_o$ are chosen so that $\ratga=0.8$.}
\begin{center}
\begin{tabular}{ccccc}
&&&&\\
$g^{\prime}_{0}$ &
$b_o  \;\;\; (m_{\pi}^{-1} )    $  &
$c_o \;\;\;  (m_{\pi}^{-3})   $  &
$M_N^*$ (MeV) &
$\ds \sqrt{\frac{<r^2>^*_p}{<r^2>_p} }$\\
&&&&\\
\hline
   1/3 &-0.024 & 0.125 & 719. & 1.089  \\
   0.6 &-0.024 & 0.150 & 714. & 1.092  \\
   0.6 &  0.0  & 0.140 & 680. & 1.10  \\
\end{tabular}
\end{center}
\label{tab2}
\end{table}

\newpage
\large
\section*{Figure captions}

\begin{description}
\item [FIG. 1.]
The profile function $\Theta (r)$ of a free skyrmion (solid curve)
and that of a skyrmion in the nuclear matter $\rho = 2.5 \rho_o$
(dashed curve). Here  $g_o^{\prime} = 0.7$.
\item [FIG. 2(a).]
The dependence of the effective nucleon mass on Lorentz -
Lorenz  parameter
$g_o^{\prime}$.
Solid, dotted and dashed curves are for $\rho = 0$, $\rho = 0.5 \rho_o$
and $\rho = \rho_o$ respectively.
\item [FIG.  2(b).]
The same as in Fig.2(a), but for the ratio $g_A^*/g_A$.
\item[FIG.  3.]
The tensor     part of the NN potential - $V_T(r)$.
Solid, dotted  and dashed
curves are for $\rho = 0$, $\rho = 0.5 \rho_o$
and $\rho = \rho_o$ respectively. Here  $g_o^{\prime} = 0.6,
b_o=-0.024 m_{\pi}^{-1}$  and $c_o=0.15m_{\pi}^{-3}$.
\end{description}
\end{document}